\documentclass[prd,english,12pt,nofootinbib,preprint]{revtex4}

\usepackage{amssymb}
\usepackage[dvips]{graphicx}

\usepackage[T1]{fontenc}
\usepackage[latin1]{inputenc}

\makeatletter

\usepackage{babel}
\makeatother

\begin{document}

\title{Higher angular momentum states of bottomonium \\ in lattice NRQCD}

\author{Randy Lewis}
\affiliation{Department of Physics and Astronomy, York University,
Toronto, Ontario, Canada M3J 1P3}

\author{R. M. Woloshyn}
\affiliation{TRIUMF, 4004 Wesbrook Mall, Vancouver, British Columbia,
Canada V6T 2A3 \\}

\begin{abstract}
On a cubic lattice the zero-momentum meson states have one of 20
possible $\Lambda^{PC}$ combinations where $\Lambda$ labels the 
irreducible representation of the octahedral group. Each continuum bottomonium
state with specific $J^{PC}$ quantum numbers is contained within one or more
of the lattice $\Lambda^{PC}$ states.
In this work, bottomonium quark-antiquark operators are constructed for 
all 20 lattice $\Lambda^{PC}$ combinations which allows many continuum
high angular momentum states to be accessible as ground states of  
their associated lattice channels.
From a dynamical simulation, realistic results are obtained for S-, P-, D-, F- 
and G-wave bottomonium states.
\end{abstract}

\maketitle

\section{Motivation}

Although the lightest spin-triplet S-wave and P-wave bottomonia have been
known experimentally for many years\cite{Nakamura:2010zzi},
the spin singlets have just been discovered in recent
experiments\cite{:2008vj,:2009pz,Bonvicini:2009hs,Adachi:2011ji}.
Beyond the P waves only the $\Upsilon(1^3D_2)$
has been observed so far\cite{Bonvicini:2004yj,delAmoSanchez:2010kz}.
Meanwhile, additional resonances, which may have an interpretation as
four-quark combinations, are being found with masses in the
bottomonium region\cite{Abe:2007tk,Belle:2011aa}.
Also, observation of the $\eta_b(2S)$ has recently been claimed, though
with a surprisingly small mass\cite{Seth,Dobbs:2012zn}.

Dynamical lattice QCD computations of the masses of
S-wave and P-wave bottomonia, including radial excitations, appear in
\cite{Gray:2005ur,Meinel:2009rd,Burch:2009az,Meinel:2010pv,
Dowdall:2011wh}. Although some of these publications also commented on
D-wave states, the first lattice study of the full set of
D-wave quantum numbers is very recent\cite{Daldrop:2011aa}.

There are many important reasons for extending lattice studies to higher
angular momentum states.  D waves and F waves provide an opportunity to confirm and
refine lattice methods, such as lattice NRQCD that uses a nonrelativistic
expansion for the valence quarks\cite{Lepage:1992tx}.
Quantitative predictions for the masses
can subsequently be tested experimentally, and comparison of reliable lattice
results with today's phenomenological models can offer insights into various
model assumptions. In particular, the spectrum as a function
of orbital angular momentum compared to radial excitations depends on the
spatial behavior of the quark-antiquark interaction. Extending the study of
bottomonium to higher partial waves augments the information gained by 
the observation of radial excited states.
The lightest G-wave bottomonium is expected to lie just below the $B\bar{B}$
threshold with F waves further below and H waves above\cite{Brambilla:2004wf},
so these bottomonium channels
offer an opportunity for using lattice QCD to study threshold issues in detail,
and to confirm which side of threshold the G waves are on.
Furthermore, knowledge of the bottomonium masses will be
crucial for understanding the four-quark candidates and any other exotic
hadrons that could be nearby.
Eventually, radiative transitions among these states could be
considered\cite{Lewis:2011ti}.

In general, a lattice operator will couple to all hadrons with the
corresponding quantum numbers, but the lightest hadron will have the cleanest
signal.  Often the lightest hadron provides the only usable signal in a lattice
study.
As a result of cubic symmetry, there are just 20 mutually orthogonal
$\Lambda^{PC}$ options available to lattice simulations.  Specifically, the
20 options originate from the reduction of continuum angular momentum
$J$ to the five lattice octahedral representations $\Lambda$, each with
$P=\pm1$ and $C=\pm1$.  All S-wave and P-wave
bottomonium states are included in that list of 20 ``ground states'',
along with just 10 higher waves.
All other continuum bottomonium quantum numbers are only
accessible as excited states in the 20 channels.
As well, any hybrids, exotics, and multiquark states would also be
included within the list of 20 possible lattice channels.

It is not known, {\em a priori}, which hadron is actually the lightest in a given
$\Lambda^{PC}$ channel.  For example, the quark model suggests that a $9^{+-}$
bottomonium state will be the lightest in one of the 20 channels,
but that same channel will couple to a $0^{+-}$ exotic hadron if one exists.
To discover which is the lighter state, it is necessary to perform an
explicit lattice simulation and analysis. Identifying the lightest mass
in each of the 20 channels is a natural starting point for lattice
QCD explorations of bottomonia at higher angular momentum.

The specific contributions of the present work are twofold.  First, 
a set of quark-antiquark creation operators for all 20 combinations of
lattice quantum numbers is constructed and the lowest continuum angular
momentum is identified.
Secondly, numerical simulations are used to show that many of
the higher angular momentum bottomonium states are accessible using present
lattice QCD methods and existing gauge field configurations.

\section{Bottomonium operators for all lattice quantum numbers}

Conservation of angular momentum is a consequence of invariance under
spatial rotations, and is realized as a tower of representations labeled by
non-negative integers $J\in\{0,1,2,\ldots\}$.
On a spatially cubic lattice, this symmetry is
reduced to the five irreducible representations
$\Lambda\in\{A_1,A_2,E,T_1,T_2\}$ of the octahedral group\cite{Johnson:1982yq}.
The well-known correspondence between $J$ and $\Lambda$ is given in
Table~\ref{tab:J2Lambda}.
\begin{table}
\caption{The number of copies of each octahedral irreducible
representation $\Lambda$, of dimension $d$,
for the subduced continuum representation $J$.}\label{tab:J2Lambda}
\begin{tabular}{lccccccccccc}
\hline\hline
$\Lambda(d)$ & \multicolumn{11}{c}{$J$} \\
       & 0 & 1 & 2 & 3 & 4 & 5 & 6 & 7 & 8 & 9 & $\cdots$ \\
\hline
$A_1(1)$ & 1 & 0 & 0 & 0 & 1 & 0 & 1 & 0 & 1 & 1 & $\cdots$ \\
$A_2(1)$ & 0 & 0 & 0 & 1 & 0 & 0 & 1 & 1 & 0 & 1 & $\cdots$ \\
$E(2)$   & 0 & 0 & 1 & 0 & 1 & 1 & 1 & 1 & 2 & 1 & $\cdots$ \\
$T_1(3)$ & 0 & 1 & 0 & 1 & 1 & 2 & 1 & 2 & 2 & 3 & $\cdots$ \\
$T_2(3)$ & 0 & 0 & 1 & 1 & 1 & 1 & 2 & 2 & 2 & 2 & $\cdots$ \\
\hline\hline
\end{tabular}
\end{table}
In principle a sophisticated analysis of excited states in lattice data
for each $\Lambda$ channel would allow the unambiguous identification of
$J$ for each observed lattice state.  However,
experience from basic quantum mechanics and the quark model suggests
that larger $J$ leads to larger bottomonium mass, so in practice it is
simple to infer each $J$ value from a single channel (unless a non quark model
hadron appears with a nearby mass).

Total angular momentum is the sum of orbital and spin components.
The orbital angular momentum $L$ of each operator corresponds to the number
of uncontracted\footnote{A lattice Laplacian, for example,
does contain a Lorentz contraction
and does not alter the orbital angular momentum.} spatial derivatives,
but it is important to remember that $\Lambda$
(the lattice version of $J$ not $L$) is the
conserved quantum number.  An operator intended for a specific $L$ will couple
to other $L$ values too, unless the fields are smeared in a way that
purposefully eliminates that mixing.

One operator for each of the 20 $\Lambda^{PC}$ channels is provided in
Table~\ref{tab:20ops}.  They are written in terms of Pauli matrices $\sigma_i$
and the symmetric lattice covariant derivative $\Delta_i$ which appears
beyond the P waves in the following forms:
\begin{eqnarray}
D_{ij} &=& \frac{1}{2}(\Delta_i\Delta_j+\Delta_j\Delta_i)
           -\frac{1}{3}\delta_{ij}(\Delta_1^2+\Delta_2^2+\Delta_3^2) \,, \\
D_{ijk} &=& \frac{1}{3!}(\Delta_i\Delta_j\Delta_k+{\rm permutations}) \,, \\
D_{ijkl} &=& \frac{1}{4!}(\Delta_i\Delta_j\Delta_k\Delta_l
             +{\rm permutations}) \,, \\
D_{ijklm} &=& \frac{1}{5!}(\Delta_i\Delta_j\Delta_k\Delta_l\Delta_m
               +{\rm permutations}) \,, \\
D_{ijklmn} &=& \frac{1}{6!}(\Delta_i\Delta_j\Delta_k\Delta_l\Delta_m\Delta_n
               +{\rm permutations}) \,, \\
D_{ijklmnopq} &=& \frac{1}{9!}(\Delta_i\Delta_j\Delta_k\Delta_l\Delta_m
                  \Delta_n\Delta_o\Delta_p\Delta_q+{\rm permutations}) \,.
\end{eqnarray}
The entries in Table~\ref{tab:20ops} can be verified by using standard
group theory methods to calculate multiplicities from the character of each
conjugacy class, but an intuitive understanding can be obtained more readily.
For example, the $^1F_3$ operator is listed as $D_{123}$. Any spatial rotation
of this operator on a lattice will reproduce the same operator except that
its indices may be permuted and for certain rotations there will be an overall
minus sign due to the vector nature of a spatial derivative.  These spatial
rotations have the properties of the one-dimensional representation $A_2$,
including the overall minus signs for the appropriate rotations.  $P$ and $C$
properties can also be verified.

\begin{table}
\caption{Creation operators $\Omega$ for each irreducible representation
$\Lambda$
of the octahedral group with specified parity $P$ and charge conjugation $C$.
Columns labeled $J^{PC}$ and $^{2S+1}L_J$ contain the
quantum numbers of quark-antiquark states obtained using the standard
relation of $S$ and $L$ with parity and charge conjugation.}\label{tab:20ops}
\begin{tabular}{llcl}
\hline\hline
$\Lambda^{PC}$ & $J^{PC}$ & $^{2S+1}L_J$ & $\Omega$ \\
\hline
$A_1^{-+}$ & $0^{-+}$ & $^1S_0$ & 1 \\
$T_1^{--}$ & $1^{--}$ & $^3S_1$ & $\{\sigma_1,\sigma_2,\sigma_3\}$ \\
$T_1^{+-}$ & $1^{+-}$ & $^1P_1$ & $\{\Delta_1,\Delta_2,\Delta_3\}$ \\
$A_1^{++}$ & $0^{++}$ & $^3P_0$ &
                        $\Delta_1\sigma_1+\Delta_2\sigma_2+\Delta_3\sigma_3$ \\
$T_1^{++}$ & $1^{++}$ & $^3P_1$ & $\{\Delta_2\sigma_3-\Delta_3\sigma_2,
\Delta_3\sigma_1-\Delta_1\sigma_3,\Delta_1\sigma_2-\Delta_2\sigma_1\}$ \\
$E^{++}$ & $2^{++}$ & $^3P_2$ & $\{(\Delta_1\sigma_1-\Delta_2\sigma_2)/\sqrt{2}
         ,(\Delta_1\sigma_1+\Delta_2\sigma_2-2\Delta_3\sigma_3)/\sqrt{6}\}$ \\
$T_2^{++}$ & $2^{++}$ & $^3P_2$ & $\{\Delta_2\sigma_3+\Delta_3\sigma_2,
      \Delta_3\sigma_1+\Delta_1\sigma_3,\Delta_1\sigma_2+\Delta_2\sigma_1\}$ \\
$E^{-+}$ & $2^{-+}$ & $^1D_2$ & $\{(D_{11}-D_{22})/\sqrt{2},
                                         (D_{11}+D_{22}-2D_{33})/\sqrt{6}\}$ \\
$T_2^{-+}$ & $2^{-+}$ & $^1D_2$ & $\{D_{23},D_{31},D_{12}\}$ \\
$E^{--}$ & $2^{--}$ & $^3D_2$ & $\{(D_{23}\sigma_1-D_{13}\sigma_2)/\sqrt{2},
                (D_{23}\sigma_1+D_{31}\sigma_2-2D_{12}\sigma_3)/\sqrt{6}\}$ \\
$T_2^{--}$ & $2^{--}$ & $^3D_2$ & $\{
                    (D_{22}-D_{33})\sigma_1+D_{13}\sigma_3-D_{12}\sigma_2,
                    (D_{33}-D_{11})\sigma_2+D_{21}\sigma_1-D_{23}\sigma_3,$ \\
        &&&        $(D_{11}-D_{22})\sigma_3+D_{32}\sigma_2-D_{31}\sigma_1\}$ \\
$A_2^{--}$ & $3^{--}$ & $^3D_3$ &
                              $D_{12}\sigma_3+D_{23}\sigma_1+D_{31}\sigma_2$ \\
$A_2^{+-}$ & $3^{+-}$ & $^1F_3$ & $D_{123}$ \\
$T_2^{+-}$ & $3^{+-}$ & $^1F_3$ & $\{
                   D_{122}-D_{133},D_{233}-D_{211},D_{311}-D_{322}\}$ \\
$A_2^{++}$ & $3^{++}$ & $^3F_3$ & $(D_{221}-D_{331})\sigma_1
                       +(D_{332}-D_{112})\sigma_2+(D_{113}-D_{223})\sigma_3$ \\
$T_1^{-+}$ & $4^{-+}$ & $^1G_4$ & $\{D_{2223}-D_{3332},D_{3331}-D_{1113},
                                     D_{1112}-D_{2221}\}$ \\
$A_1^{--}$ & $4^{--}$ & $^3G_4$ & $(D_{2223}-D_{3332})\sigma_1
                   +(D_{3331}-D_{1113})\sigma_2+(D_{1112}-D_{2221})\sigma_3$ \\
$E^{+-}$ & $5^{+-}$ & $^1H_5$ & $\{(D_{23111}-D_{13222})/\sqrt{2},
                (D_{23111}+D_{13222}-2D_{12333})/\sqrt{6}\}$ \\
$A_2^{-+}$ & $6^{-+}$ & $^1I_6$ & $D_{112222}+D_{223333}+D_{331111}
                                  -D_{221111}-D_{332222}-D_{113333}$ \\
$A_1^{+-}$ & $9^{+-}$ & $^1L_9$ & $D_{122233333}+D_{233311111}
                   +D_{311122222}-D_{133322222}-D_{211133333}-D_{322211111}$ \\
\hline\hline
\end{tabular}
\end{table}

For use with lattice NRQCD, each of the operators in Table~\ref{tab:20ops}
will be sandwiched between a bottom creation operator $\psi^\dagger(x)$
and an antibottom creation operator $\chi^\dagger(x)$, then
summed spatially to produce a zero-momentum interpolating operator:
\begin{equation}\label{operators}
\sum_{\vec x}\psi^\dagger(\vec x)\Omega\chi^\dagger(\vec x) \, .
\end{equation}
The $\psi^\dagger(x)$ and $\chi^\dagger(x)$ can contain internal summations
as well, to represent spatial smearing.
In our simulations, covariant derivatives act on the field to the right and
not to the left.  Charge conjugation is maintained exactly for all operators
because of the sum over $\vec x$ in Eq.~(\ref{operators}).

A table similar to Table~\ref{tab:20ops} appears in \cite{Davies:1994mp}, but
with a few notable differences.  In \cite{Davies:1994mp} the authors listed
only S, P and D waves, and their double derivative operator $\Delta_{ij}$ is not
symmetrized in its indices, whereas we prefer to use a symmetric definition so
exact symmetries are apparent.
For the $E^{PC}$ channels, which are two-dimensional representations, the
authors of \cite{Davies:1994mp}
chose to list three nonindependent projections whereas we prefer to
list the two independent projections.  They also included three
additional operators in the $T_1^{--}$ and $T_2^{--}$ channels,
intended to probe D-wave states which, however, are not expected to be
ground states in their respective $\Lambda^{PC}$ channels;
these operators are displayed in our
Table~\ref{tab:extraops} for convenient reference.
For relativistic operators rather than NRQCD operators, as might be
more appropriate for charmonium, see \cite{Liao:2002rj,Liu:2012ze}.
\begin{table}
\caption{Creation operators $\Omega$ that duplicate some quantum numbers from
Table~\protect\ref{tab:20ops}, but could be useful for studies of the D-wave
spectrum.  They were first listed in
\protect\cite{Davies:1994mp}.}\label{tab:extraops}
\begin{tabular}{llcl}
\hline\hline
$\Lambda^{PC}$ & $J^{PC}$ & $^{2S+1}L_J$ & $\Omega$ \\
\hline
$T_1^{--}$ & $1^{--}$ & $^3D_1$ & $\{
                      D_{11}\sigma_1+D_{12}\sigma_2+D_{13}\sigma_3,
                      D_{21}\sigma_1+D_{22}\sigma_2+D_{23}\sigma_3,
                      D_{31}\sigma_1+D_{32}\sigma_2+D_{33}\sigma_3\}$ \\
$T_1^{--}$ & $3^{--}$ & $^3D_3$ & $\{
                    3D_{11}\sigma_1-2D_{12}\sigma_2-2D_{13}\sigma_3,
                    3D_{22}\sigma_2-2D_{23}\sigma_3-2D_{21}\sigma_1,$ \\
               &&& $3D_{33}\sigma_3-2D_{31}\sigma_1-2D_{32}\sigma_2\}$ \\
$T_2^{--}$ & $3^{--}$ & $^3D_3$ & $\{
                  (D_{22}-D_{33})\sigma_1+2D_{12}\sigma_2-2D_{13}\sigma_3,
                  (D_{33}-D_{11})\sigma_2+2D_{23}\sigma_3-2D_{21}\sigma_1,$ \\
             &&& $(D_{11}-D_{22})\sigma_3+2D_{31}\sigma_1-2D_{32}\sigma_2\}$ \\
\hline\hline
\end{tabular}
\end{table}

Although the list of 20 quantum numbers $\Lambda^{PC}$ is complete,
the operators listed for each are not unique.  The operators displayed in
Table~\ref{tab:20ops} were chosen with standard bottomonium states in mind,
so each contains one quark, one antiquark and $L$ derivatives.
These operators will in general have some overlap with exotics, hybrids and
multiquark states as well, but a search meant to target those states should
construct other operators with the goal of having a larger overlap with the
particular states of interest (see, for example, \cite{Liu:2012ze}).

The $^3D_1$ meson does not appear in Table~\ref{tab:20ops} because it has
the same $J^{PC}$ as the lighter $^3S_1$ meson.  Similarly,
the $^3F_2$ meson does not appear in Table~\ref{tab:20ops} because it has
the same $J^{PC}$ as the lighter $^3P_2$ meson.  Neither of those omissions
is a consequence of lattice discretization, since even the continuum
quantum numbers are shared.  On the other hand, the absence of the $^3F_4$
meson from Table~\ref{tab:20ops} is a consequence of lattice discretization:
its continuum quantum numbers ($4^{++}$) are not shared by any lighter state,
but its lattice quantum numbers ($A_1^{++}$, $E^{++}$, $T_1^{++}$, $T_2^{++}$)
are all shared by the $^3P_J$ mesons.  Therefore
lattice signals of the $^3D_1$, $^3F_2$, $^3F_4$ and many similar cases must be
sought as excited states in the corresponding $\Lambda^{PC}$ channels,
unless operators can be tuned carefully enough to remove any coupling to
the lighter mesons in each channel.
Analysis of excited states in any $\Lambda^{PC}$ channel
is beyond the scope of the present work with one exception: we will
address the $\eta_b(2S)$ because of the present experimental
situation\cite{Seth,Dobbs:2012zn}.

\section{Lattice simulations}

All numerical simulations performed for this work utilize a 2+1-flavor
dynamical gauge field ensemble
generated by the PACS-CS Collaboration\cite{Aoki:2008sm}
and made available through the Japan Lattice Data Grid\cite{JLDG}
as part of the International Lattice Data Grid\cite{Beckett:2009cb}.
This ensemble uses the Iwasaki gauge action and the nonperturbatively tuned
clover action for up, down, and strange quarks. The light quark (up and down)
masses are such that the pion mass is 156 MeV$/c^2$, i.e.\ near the
physical value. There are 198 configurations with
lattice volume $32^3\times64$.  The four input parameters, along with
four basic output quantities, are displayed in Table~\ref{tab:cfgs}.
Not provided by the PACS-CS Collaboration
but needed for the present work is the mean link 
in Landau gauge that was computed\cite{Lewis:2011ti} to be $u_L = 0.8463$.
\begin{table}
\caption{Input parameters for the ensemble of gauge configurations generated
by the PACS-CS Collaboration\protect\cite{Aoki:2008sm}.
Some standard output quantities provided by PACS-CS are also
listed.}\label{tab:cfgs}
\begin{tabular}{ll}
\hline\hline
Inputs & Standard outputs \\
\hline
$\beta=1.90$, & $a=0.0907(13)$ fm \\
$\kappa_{ud}=0.13781$, & $m_\pi=156(7)$ MeV$/c^2$ \\
$\kappa_s=0.13640$, & $m_K=554(8)$ MeV$/c^2$ \\
$c_{SW}=1.715$, & $\langle\square\rangle=0.570870(5)$ \\
\hline\hline
\end{tabular}
\end{table}

The bottom quark and antiquark are described by the lattice NRQCD
Hamiltonian\cite{Lepage:1992tx} omitting terms at $O(v^6)$ and beyond.
Specifically, the Hamiltonian from the Appendix of \cite{Lewis:2008fu} is
used with $c_i=1$ for $i\leq6$ and $c_i=0$ for $i\geq7$, with tadpole
factor $U_0=u_L=0.8463$ and with stability parameter $n=4$.

In previous work\cite{Lewis:2011ti} the bottom quark bare mass was tuned to
match the experimental $\eta_b(1S)$ mass, with result $M_b=1.945$.  However,
a recent measurement by the Belle Collaboration\cite{Collaboration:2011chb}
is significantly larger than the $\eta_b(1S)$ mass reported by previous
experiments\cite{Nakamura:2010zzi}.  Tuning to the Belle result would require
a slight increase in the bottom quark bare mass of lattice NRQCD.  To avoid
this unresolved issue, it is preferable to use the well-established
$\Upsilon(1S)$ mass\cite{Nakamura:2010zzi}
for tuning instead.  This also leads to a slight increase in the bare mass
parameter relative to our previous work\cite{Lewis:2011ti}, and for the present
project we use $M_b=1.95$.

To reduce statistical fluctuations, it would be optimal to average over
results obtained from putting the creation operator at each lattice site.
A practical implementation of this general idea is employed here:
heavy quark propagators are constructed from a random-U(1) wall
source\cite{Davies:2007vb} spanning all spatial lattice sites at a
fixed Euclidean time.  A further reduction of statistical fluctuations can be
achieved by selecting another Euclidean time on the same gauge configuration,
separately inserting a random-U(1) wall source there,
and averaging the two correlation functions on that configuration.
For the present work, averages are taken over 64 separate calculations of
all correlation functions per configuration, corresponding to a random-U(1)
wall source at each Euclidean time on the $32^3\times64$ lattice.

Excellent ground state signals are obtained for S and P waves with 
a single correlation function using a local operator.
As one goes to higher waves, simulation energies increase and correlation 
functions decrease more rapidly and become noisier due to the presence of 
more derivatives in the operators. It is advantageous to utilize more
correlators in each channel and to smear the operators to enhance the 
ground state contribution at smaller time separations.
Beyond the P waves, gauge-invariant smearing\cite{Alford:1995dm} is applied to
the bottom quark operator,
\begin{equation}
\psi(x) \to \left(1+0.15\Delta^2\right)^{8s}\psi(x) \, .
\end{equation}
Three options are considered: $s=0$ (local), $s=1$ (smeared), and $s=2$
(doubly smeared).
In every case the antiquark field $\chi(x)$ remains unsmeared.

With local, smeared and doubly smeared operators available for both creation
($s$) and annihilation ($s^\prime$),
each $\Lambda^{PC}$ has nine correlation functions to be
analyzed simultaneously.  The nominal fit function is
\begin{equation}\label{nominalfit}
g_{s,s^\prime}(t-t_0) = \sum_{n=1}^Nf_s(n)f_{s^\prime}(n)e^{-E_n(t-t_0)}
\end{equation}
where the number of exponentials $N$ is increased until the $\chi^2$/d.o.f.\
is unity or less.  $E_n$ denotes the simulation energy for the $n$th meson
while $f_s(n)$ and $f_{s^\prime}(n)$ denote the coupling of the $n$th meson to
the creation and annihilation operators respectively.

In practice, the lattice data cannot always resolve
individual higher-energy states and that requires a generalization of
Eq.~(\ref{nominalfit}).  As a simple example consider two operators, $A$ and $B$,
that each couple to two physical mesons having energies $E_1$ and $E_2$.
Suppose these energies are unequal but not distinguishable with the numerical
precision of the available lattice data, so they get represented
by a single exponential during the fitting process,
$E_1\approx E_2\approx E_{\rm fit}$.  The nominal fit function
(with $t_0=0$) would give
\begin{eqnarray}
\langle0|AA^\dagger|0\rangle &=& \left[f_A(1)\right]^2e^{-E_1t}
                               + \left[f_A(2)\right]^2e^{-E_2t}
~\approx~ \bigg(\left[f_A(1)\right]^2+\left[f_A(2)\right]^2\bigg)
            e^{-E_{\rm fit}t} , \\
\langle0|BB^\dagger|0\rangle &=& \left[f_B(1)\right]^2e^{-E_1t}
                               + \left[f_B(2)\right]^2e^{-E_2t}
~\approx~ \bigg(\left[f_B(1)\right]^2+\left[f_B(2)\right]^2\bigg)
           e^{-E_{\rm fit}t} , \\
\langle0|BA^\dagger|0\rangle &=& f_A(1)f_B(1)e^{-E_1t}
                               + f_A(2)f_B(2)e^{-E_2t}
~\approx~ \bigg(f_A(1)f_B(1)+f_A(2)f_B(2)\bigg)e^{-E_{\rm fit}t} .~~~~~
\end{eqnarray}
The three coefficients of $e^{-E_1t}$ contain only two degrees of freedom
as assumed in Eq.~(\ref{nominalfit}).  The same is true for the coefficients of
$e^{-E_2t}$.  However, the three coefficients of $e^{-E_{\rm fit}t}$ do not
contain
just two degrees of freedom.  To account for this potential inability of
lattice data to resolve excited states, Eq.~(\ref{nominalfit}) can be
generalized to
\begin{equation}\label{generalfit}
g_{s,s^\prime}(t-t_0)
         = \sum_{n=1}^{N^\prime}f_s(n)f_{s^\prime}(n)e^{-E_n(t-t_0)}
         + \sum_{n=N^\prime+1}^Nf_{s,s^\prime}(n)e^{-E_n(t-t_0)}
\end{equation}
which says the lowest $N^\prime$ states are resolved individually but
higher states may not be.
Confident extraction of the ground-state energy requires $N^\prime>1$,
which is satisfied easily by all of the data discussed in this paper.

The methods described above are sufficient to permit explorations in the
S, P, D, F, G, H and I channels as presented in the following section, with
reasonably precise results obtained in the S, P and D channels.
Since F and G waves have not been discussed in previous lattice simulations
but are expected to lie below the $B\bar{B}$
threshold\cite{Brambilla:2004wf},
we focus particular attention on the F- and G-wave channels by implementing,
in addition to the methods described above,
stout spatial links $\widetilde{U}$ as proposed in \cite{Morningstar:2003gk}:
\begin{eqnarray}
&& U \to U^{(1)} \to U^{(2)} \to \cdots \to U^{(n_\rho)} \equiv \widetilde{U}
\,, \\
&& U_\mu^{(n+1)}(x) = \exp\left(iQ_\mu^{(n)}(x)\right)
   U_\mu^{(n)}(x) \,,
\end{eqnarray}
where $Q_\mu^{(n)}(x)$ contains a product of gauge
links weighted by a parameter
$\rho$.  [For the precise definition of $Q_\mu^{(n)}(x)$, see
\cite{Morningstar:2003gk}.]  Our F- and G-wave simulations use $\rho=0.2$ and
$n_\rho=16$ for source and sink operators, but no stouting within the NRQCD
propagator.

Lattice simulations involve both systematic and statistical errors.
In the present context, systematic errors arise from the neglect of $O(v^6)$
terms in the NRQCD Hamiltonian, the neglect of $O(\alpha_s)$ corrections to the
NRQCD coefficients $c_i$, the nonzero lattice spacing, the finite lattice
volume, and up and down quark masses that are slightly larger than their
physical values.  Although we do not have any quantitative means to determine
systematic errors for ourselves, the authors of \cite{Gray:2005ur} studied
bottomonium with the same truncation of NRQCD terms and tadpole-improved
tree-level $c_i$ definitions on lattices having discretizations and volumes
similar to ours, though their up and down quarks were significantly heavier
than ours.  The thorough discussion in \cite{Gray:2005ur} concluded that
systematic errors in typical $\Upsilon$ mass splittings are at the few-percent
level.  We take this as a generic expectation for our systematic errors too,
though only statistical errors will be reported below.  Interestingly, the
statistical errors will turn out to be at the few-percent level also.

\section{Lattice results}

\begin{figure}
\includegraphics[width=10cm,angle=270,trim=80 50 50 90,clip=true]{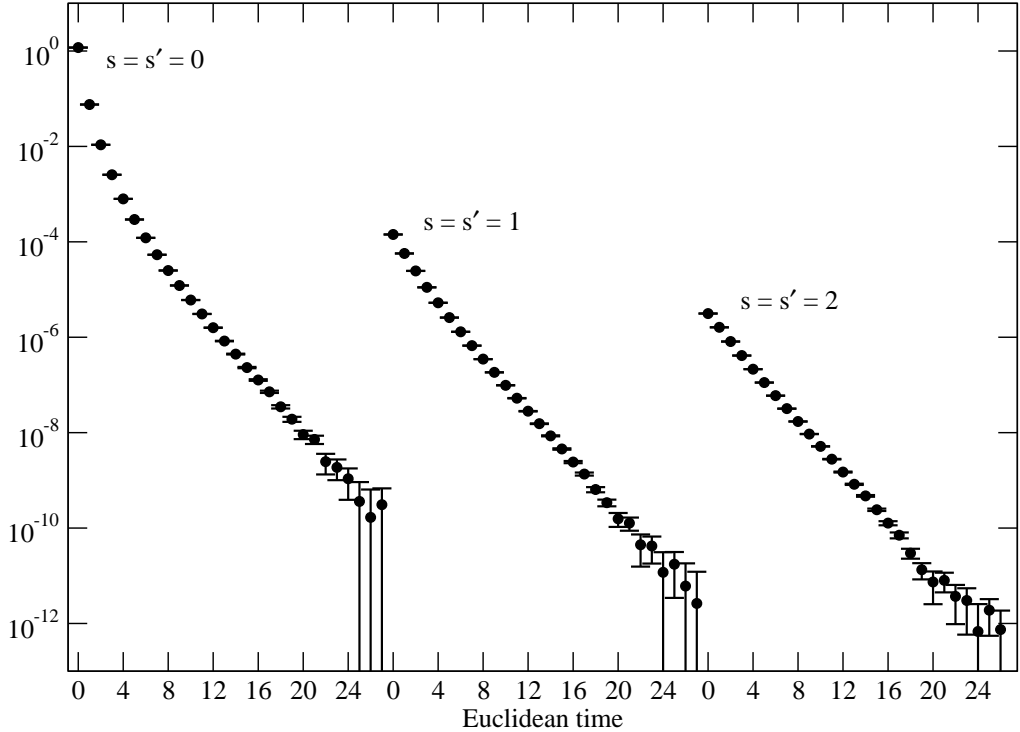}
\caption{Correlation functions for $E^{--}$ with equal smearing at source and
         sink.  The lightest meson in this channel is $^3D_2$
         bottomonium.}\label{fig:Dwave}
\end{figure}

\begin{figure}
\includegraphics[width=10cm,angle=270,trim=80 50 50 90,clip=true]{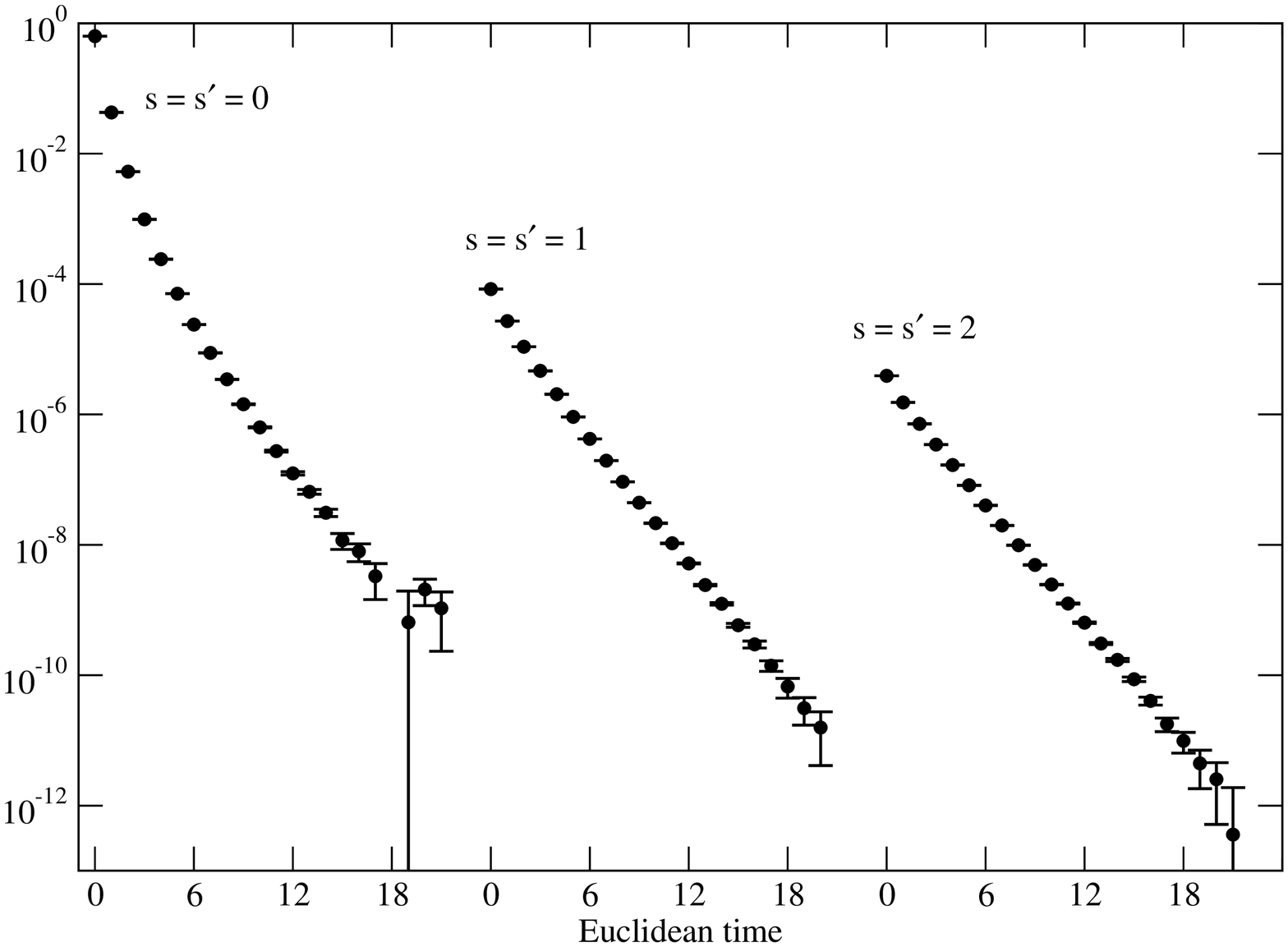}
\caption{Correlation functions for $T_2^{+-}$ with equal smearing at source
         sink.  Stout links are used at source and
         sink as well.  The lightest meson in this channel is
         $^1F_3$ bottomonium.}\label{fig:Fwave}
\end{figure}

To provide a visual impression of the quality of lattice data, sample
D-wave and F-wave correlation functions are displayed in
Figs.~\ref{fig:Dwave} and \ref{fig:Fwave} respectively.
Numerical results for the fits are listed in Table~\ref{tab:energies}.
Because the smearing parameters were not optimized individually for each
operator, two of the fits do not benefit from using all nine combinations of
source and sink smearings; Table~\ref{tab:energies} records the number of
combinations used in each case.
Fits employ the standard correlated $\chi^2$ method\cite{Gattringer:2010zz}.
With a source at Euclidean time $t_0$, the S- and P-wave fits include
$1\leq t-t_0\leq31$, D-wave fits use $1\leq t-t_0\leq27$, and all higher-wave
fits use
$1\leq t-t_0\leq23$. The conversion from the lattice simulation energy
$E_{\rm sim}$ to a meson mass in physical units (also shown in
Table~\ref{tab:energies}) is accomplished by recalling that $M_b$ was tuned
to the physical $\Upsilon(1S)$ mass.  Explicitly,
\begin{equation}
Mc^2 = M(\Upsilon(1S)_{\rm experiment})c^2
    + \frac{\hbar c}{a}\bigg(E_{\rm sim}-E_{\rm sim}(\Upsilon(1S))\bigg)
\end{equation}
where the lattice spacing, $a$, is taken from Table~\ref{tab:cfgs}.
\begin{table}
\caption{Simulation energies and resulting bottomonium masses obtained from
fits to lattice
data.  Also shown are the number of correlators (i.e.\ smearing combinations)
used in the fits and the number of exponentials needed to reduce the
$\chi^2/$d.o.f.\ to unity or less.  All operators are from
Table~\protect\ref{tab:20ops} except the one labeled $T_2^{--}(^3D_3)$
which is from Table~\protect\ref{tab:extraops}.  Errors are statistical
only.}\label{tab:energies}
\begin{tabular}{ccclc}
\hline\hline
$\Lambda^{PC}$ & $n_{\rm corr}$ & $n_{\rm exp}$ & $E_{\rm sim}$
 & mass [GeV/$c^2$] \\
\hline
$A_1^{-+}$ & 1 & 4 & 0.23420(6) & 9.4040(8) \\
$T_1^{--}$ & 1 & 4 & 0.25993(7) & 9.460 \\
$T_1^{+-}$ & 1 & 4 & 0.451(4) & 9.876(11) \\
$A_1^{++}$ & 1 & 4 & 0.438(4) & 9.847(11) \\
$T_1^{++}$ & 1 & 4 & 0.449(4) & 9.871(11) \\
$E^{++}$   & 1 & 4 & 0.457(7) & 9.889(17) \\
$T_2^{++}$ & 1 & 4 & 0.454(7) & 9.882(17) \\
$E^{-+}$   & 9 & 5 & 0.591(5) & 10.180(16)\\
$T_2^{-+}$ & 9 & 7 & 0.599(4) & 10.198(14) \\
$E^{--}$   & 9 & 6 & 0.599(4) & 10.198(14) \\
$T_2^{--}$ & 9 & 6 & 0.588(3) & 10.174(13) \\
$A_2^{--}$ & 9 & 6 & 0.600(3) & 10.200(13) \\
$T_2^{--}(^3D_3)$ & 9 & 6 & 0.601(3) & 10.202(13) \\
$T_2^{+-}$ & 9 & 6 & 0.684(4) & 10.369(17) \\
$A_2^{+-}$ & 9 & 5 & 0.690(3) & 10.383(16) \\
$A_2^{++}$ & 9 & 5 & 0.683(5) & 10.367(18) \\
$T_1^{-+}$ & 6 & 4 & 0.775(1) & 10.581(17) \\
$A_1^{--}$ & 6 & 4 & 0.778(3) & 10.587(18) \\
\hline\hline
\end{tabular}
\end{table}

The $\eta_b(1S)$ mass in Table~\ref{tab:energies} is consistent
with the Belle measurement and larger than the PDG average:
\begin{equation}
M(\eta_b(1S)) = \left\{\begin{array}{ll}
            9390.9\pm2.8 {\rm ~MeV}/c^2, &
                                  \mbox{{\rm PDG~average}\cite{Nakamura:2010zzi}} \\
            9401.0\pm1.9^{+1.4}_{-2.4} {\rm ~MeV}/c^2, &
                                   \mbox{{\rm Belle}\cite{Collaboration:2011chb}} \\
            9404.0\pm0.8 {\rm ~MeV}/c^2, &
                                  {\rm this~work,~statistical~errors~only.}
            \end{array}\right.
\end{equation}
It should be noted that this lattice result is consistent with some lattice
determinations that predated the
Belle measurement\cite{Gray:2005ur,Meinel:2009rd,Lewis:2011ti}.
Studies of specific aspects of this issue are also available in
\cite{Meinel:2010pv,Hammant:2011bt}.

The P-wave and D-wave states are
in reasonable agreement with the available experimental measurements,
but splittings among D waves are not resolved, in contrast to
\cite{Daldrop:2011aa}.
The three F-wave results from Table~\ref{tab:energies},
\begin{eqnarray}
M(^1F_3) &=& \left\{\begin{array}{l} 10.369\pm0.017 {\rm ~GeV}/c^2 \, , \\
                                     10.383\pm0.016 {\rm ~GeV}/c^2 \, , \end{array}
             \right. \\
M(^3F_3) &=& 10.367\pm0.018 {\rm ~GeV}/c^2 \, ,
\end{eqnarray}
are in excellent agreement with a model prediction of
10.37 GeV$/c^2$\cite{Brambilla:2004wf}.
The two G-wave results from Table~\ref{tab:energies},
\begin{eqnarray}
M(^1G_4) &=& 10.581\pm0.017 {\rm ~GeV}/c^2 \, \\
M(^3G_4) &=& 10.587\pm0.018 {\rm ~GeV}/c^2 \, ,
\end{eqnarray}
are about one standard deviation above 
the experimental $B\bar{B}$ threshold at 10.56 GeV$/c^2$.
Recalling that the quoted lattice uncertainties are statistical only,
our results are also reasonably close to the model prediction of
10.52 GeV$/c^2$\cite{Brambilla:2004wf} (which, however, lies below the 
$B\bar{B}$ threshold). 

Fits to lattice data in H-wave and I-wave channels could only support three or
four exponentials, had no stout links, and lay above the $B\bar{B}$
threshold, so meaningful mass predictions were not obtained.
We did not attempt an L-wave simulation.

\begin{figure}
\includegraphics[width=12cm,angle=270,trim=0 0 0 0,clip=true]{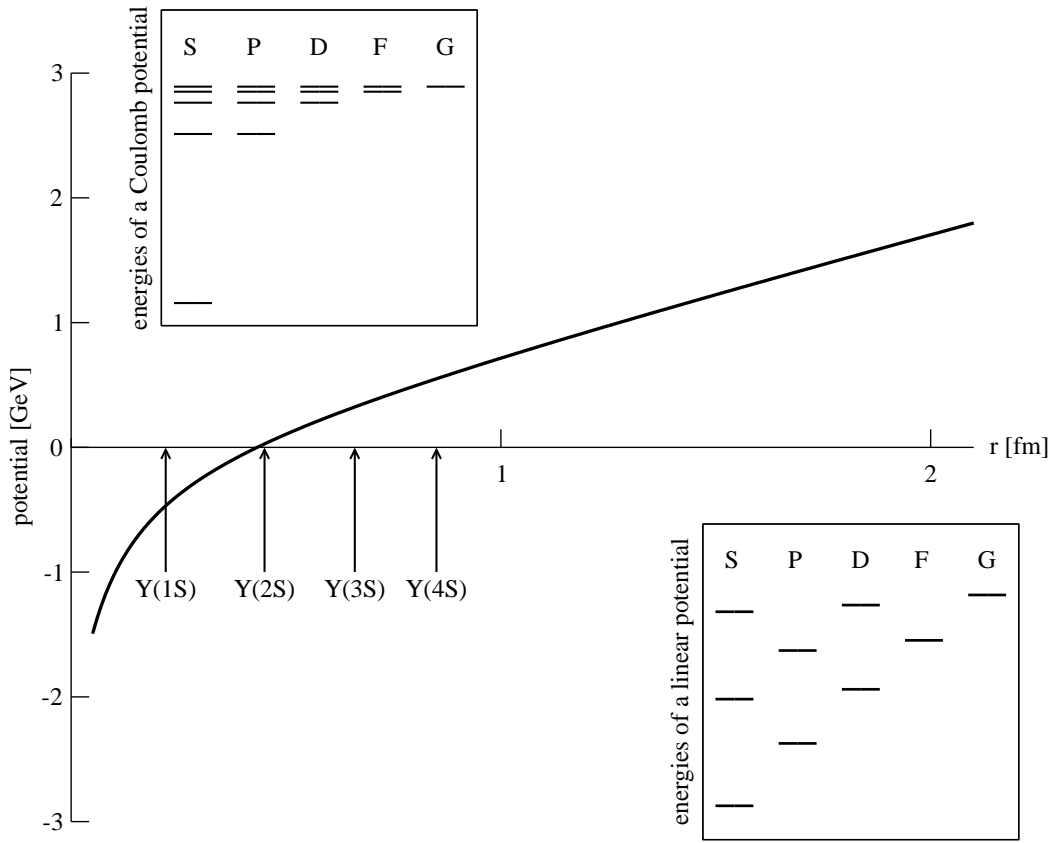}
\caption{{\bf Main graph:}
         The $q\bar{q}$ potential for mesons in the relativized quark model of
         Godfrey and Isgur\cite{Godfrey:1985xj}; see their Fig.~12.
         {\bf Upper inset:} 
         The spectrum of low-lying energy levels from a simple Coulomb
         potential.
         {\bf Lower inset:} 
         The spectrum of low-lying energy levels from a simple linear
         potential.
         }\label{fig:CoulombLinear}
\end{figure}
\begin{figure}
\includegraphics[width=10cm,angle=270,trim=80 25 50 80,clip=true]{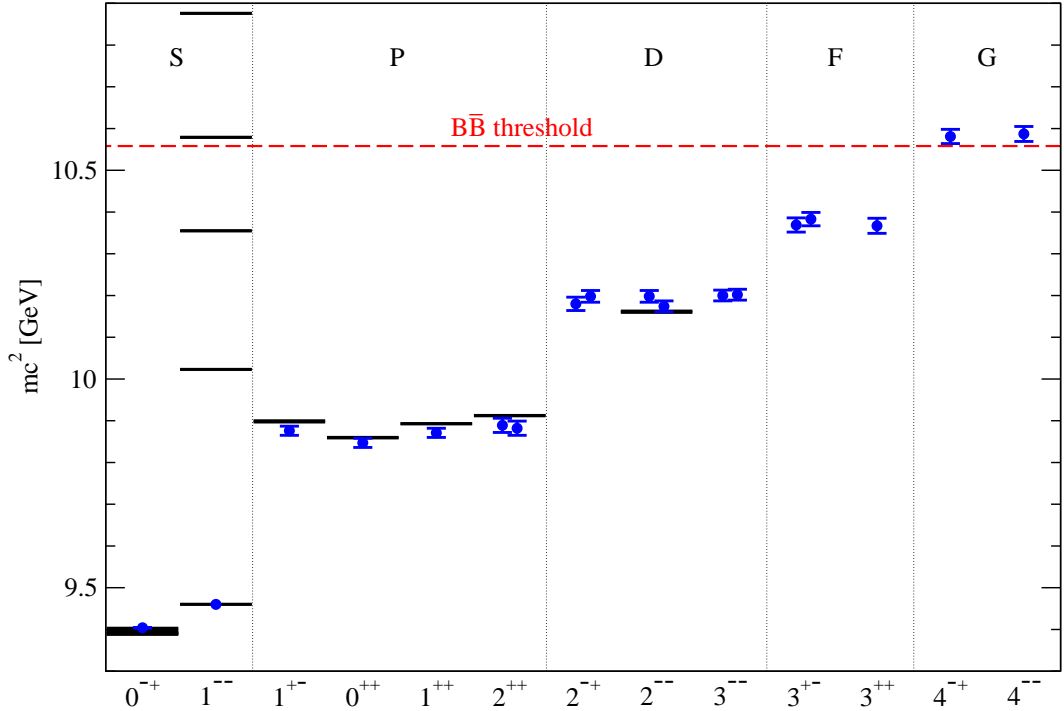}
\caption{Lattice results for the bottomonium masses (data points with
         statistical error bars) compared with experimental
         data\cite{Nakamura:2010zzi} (horizontal
         lines).  Radially excited $\Upsilon$ masses are shown for comparison
         to orbital excitations.}\label{fig:spectrum}
\end{figure}

The behavior of orbital excitations compared to radial excitations depends
on the form of the potential.
Figure~\ref{fig:CoulombLinear} shows the $q\bar{q}$
potential used in the quark model of Godfrey and Isgur\cite{Godfrey:1985xj},
along with their determination of
the location of some $\Upsilon$ excitations in that potential.
The $\Upsilon(1S)$ is in the Coulomb regime but higher radial excitations
move into the transition region between Coulomb and linear.  The insets in
Fig.~\ref{fig:CoulombLinear} show radial and orbital excitations in the
simple cases of purely Coulomb and purely linear potentials. In calculations
and from experimental data one finds that the ground-state P-wave states
have masses not very different from the first S-wave radial excitations.
There is at least some resemblance to the expectation from a Coulomb-like
potential. The situation changes as angular momentum is increased. At 
G waves the ground state would be comparable to the $\Upsilon(5S)$ in the 
Coulomb limit but
comparable to $\Upsilon(3S)$ in the linear limit.
The lattice spectrum from the present work is displayed in
Fig.~\ref{fig:spectrum}, where the G wave is seen to be comparable to the
$\Upsilon(4S)$ mass, i.e.\ intermediate to the Coulomb and linear cases.
For the lattice results, the $G-F$ splitting is comparable to the $F-D$
splitting and thus more reminiscent of a linear-dominated potential than a
Coulomb-dominated potential. 
Perhaps the most important conclusion about G waves in this work is that
statistical errors are small enough to warrant future systematic lattice
study to determine, for example, whether the G waves lie above or below 
threshold.

Although this work has focused exclusively on the ground state masses
for each $\Lambda^{PC}$, excited state masses were obtained during the
fitting process.  In light of the present experimental situation\cite{Seth,Dobbs:2012zn},
the $\eta_b(2S)$ is of particular interest.  Our present work relies on
local operators for the S waves, but our previous work\cite{Lewis:2011ti} 
included smearing and thus produced a more precise result for the $\eta_b(2S)$
mass.  Since it was
not reported explicitly in that publication, we include it here:
\begin{equation}
M\left(\Upsilon(2S)\right) - M\left(\eta_b(2S)\right)
= 24\pm3 {\rm ~MeV}/c^2
\end{equation}
where the error is statistical only.
This hyperfine splitting is significantly smaller than the recently announced
experimental measurement\cite{Seth,Dobbs:2012zn},
$M\left(\Upsilon(2S)\right) - M\left(\eta_b(2S)\right) = 48.7\pm2.7$ MeV$/c^2$,
but our result is consistent with other lattice
predictions\cite{Gray:2005ur,Burch:2009az,Meinel:2010pv,Dowdall:2011wh} as tabulated
in \cite{Seth}.

\section{Conclusions}

The study of bottomonium at increasing angular momentum provides knowledge 
that augments that obtained from the spectrum of radial excitations. In this 
paper we provide a set of operators in lattice NRQCD for higher angular 
momentum bottomonium states and explore the feasibility of their use in
numerical simulations.

A set of quark-antiquark operators for all lattice irreducible 
representations, $\Lambda^{PC}$, has been given in Table~\ref{tab:20ops}.  
In principle, this set of operators allows for bottomonium
states with 16 distinct continuum $J^{PC}$ combinations to be 
described as ground states in their respective lattice channels.
All the S- and P-wave states are in this set. However, for higher
angular momentum, some continuum $J^{PC}$ combinations 
would only appear in $\Lambda^{PC}$ channels where a lower angular 
momentum state is present. These states would have to be extracted
from lattice simulations as excited states.
Since ground state energies are much easier to determine from 
lattice simulations, the states described by the operators in
Table~\ref{tab:20ops}
are a natural starting point for numerical studies of 
the high angular momentum states of bottomonium.

Our numerical
simulations produced clear signals for S-, P-, D-, F- and G-wave states as displayed
in Fig.~\ref{fig:spectrum}.  These are the first lattice simulations of
F-wave and G-wave bottomonium, and they are near quark model expectations.
The G waves are very close to the physical $B\bar{B}$ threshold, and
statistical errors in the lattice simulation are small enough to encourage
future studies of the systematic errors so that lattice methods can
ultimately predict with confidence whether the G waves lie above or below
threshold.
Table~\ref{tab:20ops} contains ground-state operators for H, I and L waves
also, though these will be significantly above the $B\bar{B}$ threshold.

\acknowledgments
We are grateful to the PACS-CS Collaboration for making their dynamical
gauge field configurations available.
R.L. thanks the ECT* in Trento for support, and the organizers of
``Beautiful Mesons and Baryons on the Lattice'' for the opportunity to
participate while this research was in its final stages.
The work was supported in part by the Natural Sciences and 
Engineering Research Council (NSERC) of Canada, and by computing resources of
the Shared Hierarchical Academic Research Computing Network
(SHARCNET)\cite{sharcnet}.

%%%%%%%%%%%%%%%%%%%%%%%%%%%

\end{document}